# Prediction of Topological Crystalline Insulator and Topological Phase Transitions in Two-dimensional PbTe Films

Yi-zhen Jia,[1] Wei-xiao Ji,[1] Chang-wen Zhang,[1*] Ping Li,[1] Shu-feng Zhang,[1] Pei-ji Wang,[1] Sheng-shi Li,[2] and Shi-shen Yan[2]

[1] School of Physics and Technology, University of Jinan, Jinan, Shandong, 250022, People's Republic of China

[2] School of Physics, State Key laboratory of Crystal Materials, Shandong University, Jinan, Shandong 250100, People's Republic of China

Topological phases, especially topological crystalline insulators (TCIs), have been intensively explored observed experimentally in three-dimensional (3D) materials. However, the two-dimensional (2D) films are explored much less than 3D TCI, and even 2D topological insulators. Based on *ab initio* calculations, here we investigate the electronic and topological properties of 2D PbTe(001) few-layers. The monolayer and trilayer PbTe are both intrinsic 2D TCIs with a large band gap reaching 0.27 eV, indicating a high possibility for room-temperature observation of quantized conductance. The origin of TCI phase can be attributed to the $p_{x,y}$-$p_z$ band inversion, which is determined by the competitions of orbital hybridization and quantum confinement effect. We also observe a semimetal-TCI-normal insulator transition under biaxial strains, whereas a uniaxial strains lead to $Z_2$ nontrivial states. Especially, the TCI phase of PbTe monolayer remains when epitaxial grow on NaI semiconductor substrate. Our findings on the controllable quantum states with sizable band gaps present an ideal platform for realizing future topological quantum devices with ultralow dissipation.

**Keywords:** Topological crystalline insulator; Band inversion; Spin-orbital coupling; First-principles calculations

The great challenges in condensed matter physics and materials science are to tune electron conduction working at room temperature with low dissipation. The appearance of band topology opens the door to the field of topological quantum states (TQSs) [1,2], such as topological Dirac semimetal [3,4], Weyl semimetal [5,6], node-line semimetal [7-8], and topological insulators (TIs) [9-15]. Topological crystalline insulators (TCIs) [16], whose metallic boundary states are protected by crystal symmetry, rather than time-reversal symmetry (TRS), are rather exotic topological phases of matter beyond the prototypical $Z_2$ TIs [17,18]. Though some 3D TCIs have been explored depending on different crystal symmetry [19-23], those relying on mirror symmetry [24] are of particular interest, as they have been experimentally observed in IV-VI SnTe, Pb$_{1-x}$Sn$_x$Te, and Pb$_{1-x}$Sn$_x$Se [25-29]. These TCI phases exhibit a variety of novel topological behaviors such as Dirac mass generation *via* ferroelectric distortion and flat band superconductivity, [30,31] leading to wide-range interests in quantum devices in spintronics.

Remarkably, TCI phases have also been extended to 2D films, such as SnTe, PbPo and graphene multilayers [32-35]. In comparison to 3D counterpart, 2D TCI phase disappears due to the mirror symmetry breaking when epitaxial grow on a substrate, which is the main reason why these TQSs are rarely established in experiments [36]. To realize practical applications, it is essential to engineer electronic band structures of 2D TCI with the desired feature. Notably, the band structure of 3D bulk PbTe has been found to be trivial, but its free-standing (001) monolayer turns into a 2D TCI [32, 33]. However, the dependence of nontrivial topology on the number of PbTe layers has not been explored, which is further limited to practical applications. To give a better understanding of TQSs, a thorough exploration of electronic and topological properties of PbTe(001) few-layers is of rather importance.

In this contribution, we perform density-functional theory (DFT) calculations to predict that the (001) monolayer and trilayer are both intrinsic 2D TCIs with band gaps reaching 0.27 eV, which make it viable for practical realization at room-temperature. This new TQS has a different symmetry class as compared with $Z_2$ TI, thus its edge states are protected solely by crystal symmetry instead of TRS. Another prominent feature is that a semimetal-TCI-normal insulator transition occurs with biaxial strain, while the uniaxial strain leads to $Z_2$ nontrivial state. Our findings enrich the family of 2D TCI phases and thus can serve as an ideal platform for observing robust TQSs experimentally.

All the calculations were performed within the framework of density-functional theory (DFT) [37] as implemented in the Vienna *ab initio* simulation package [38]. We employ the Perdew-Burke-Ernzerhof (PBE) exchange-correlation functional of the projector augmented-wave (PAW) potential with the $6s^2 6p^2$ state of Pb and $4d^{10}5s^25p^4$ of Te as valence electrons, [39-41]. The plane-wave basis set with energy cutoff of 500 eV and Γ-centered Monkhorst–Pack k-mesh of (15×15×1) was used. The vacuum space was set to 20 Å to minimize artificial interactions between neighboring slabs. Considering the possible underestimation of the band gap with PBE, the nonlocal HSE06 hybrid functional is further supplemented to check the band topology of PbTe few-layers. [42-43] During the structural optimization, the maximum force allowed on each atom was less than 0.02 eV/ Å. Spin-orbit coupling (SOC) was included by a second vibrational procedure on a fully self-consistent basis.

Bulk PbTe has a face-centered cubic NaCl-type structure with a narrow band gap at the *L* point, forming a 3D trivial insulator, which has multiple applications in the thermoelectricity, infrared diode, and even superconductivity [24,44,45]. When it is grown along (001) direction in a layer-by-layer model, the (001) few-layers with a square Bravais lattice form a symmetry of point group $C_4$, as illustrated in Fig. 1(a). Obviously, this structure is different from the hexagonal homeycomb 2D materials. [46-49]. To get few-layer films, we cut the bulk PbTe along (001) direction, thus the even-layers comprise of a combination of bilayer blocks, whereas the odd-layers have mirror symmetry along *z* direction. Hence the time-reversal invariant momentum (TRIM) points locate at Γ, M, and X(Y) points, as illustrated in Fig. 1(b). We calculate the binding energies of PbTe few-layers expressed as $E_b$ = E(PbTe)$_n$ – $n$E(Pb) – $n$E(Te), where E (PbTe), E (Pb), and E (Te) are the total energies of PbTe few-layers, bulk Pb and Te crystals, respectively. The calculated results are -4.14, -4.52, and -4.71 eV for single-layer, bilayer and trilayer, respectively. In comparison to the bulk counterpart, structural optimizations indicate the lattice constant of (001) few-layers decrease drastically, but the intra-layer atomic distance changes slightly. As such, the distances between PbTe sheets in trilayer PbTe increases by 0.9 %. Furthermore, we examine the phonon spectra of the single-layer and bilayer PbTe, as shown in Figs. S1 and S2 in the supplementary information. Note that the monolayer structure is dynamical unstable. However, experimentally, the 2D free-standing films should be placed or grown on a substrate, thus our calculated electronic properties of monolayer PbTe is still interesting.

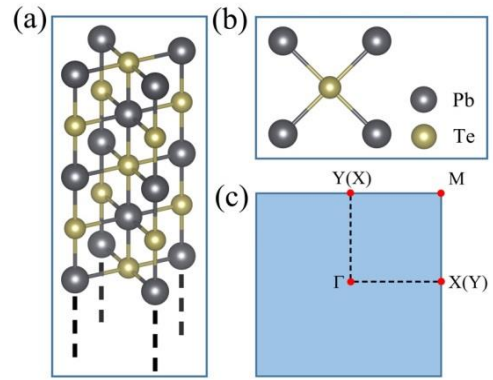

**Fig. 1** (a) The side and (b) top views of crystal structure of (001) few-layers with the unit cell indicated by blue box. (c) 2D Brillouin zones with specific symmetry points labeled.

Figure 2 shows the band structures of (001) few-layers calculated from monolayer to 6-layer PbTe (001). It is noted that the band gap gradually decreases as the layer number increases, though it oscillates between odd and even layers. This scenario is in analogy to the case of 2D PbS [50], where the band gaps have also variant oscillatory. However, when the layer thickness reaches eight, the band gap remains constant, closer to the experimental bulk value [32]. In fact, the band gaps vary from 0.11 to 1.22 eV for all (001) few-layers, which enables these few-layers promising candidates for nanoelectronics devices.

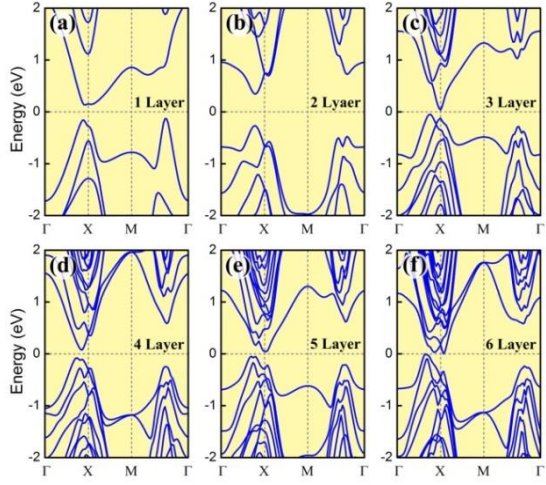

**Fig. 2** (a-f) The calculated band structures of (001) PbTe thin films from monolayer to hexalayer respectively.

We now turn to the orbital character of band structure near the X point with an odd number of atomic layers, due to the mirror symmetry under the reflection of $z \rightarrow -z$. By examining orbital-projected electronic states, we find an interesting layer-dependent band order, *i.e.*, both the monolayer and trilayer crystals have a clear band inversion, as shown in Fig. 3. Take the (001) monolayer as an example, in the absence of SOC, the valence band maximum (VBM) at the X point is derived mainly from Te-$5p_{x,y}$ with positive (+) parity, while the conduction band minimum (CBM) originates predominantly from Pb-$6p_z$ with negative (-) parity. Along with turning on SOC, the band components of Te-$5p_{x,y}$ and Pb-$6p_z$ states at the band edge of X the point are interchanged with each other, see Fig. 3(a). Such a fact demonstrates the existence of band inversions in (001) monolayer, which is signature of the existence of the TCI phase. Similar phenomena have been obtained for trilayer PbTe, as illustrated in Fig. 3(b), although the band gap decreases as compared with the monolayer one. However, when the number of layers increases beyond the penetration length of surface states, *i.e.*, a critical number of tetralayer, the weak orbital hybridization between two surfaces would lead to a small energy splitting between the bonding Te-$5p_{x,y}$ and anti-bonding Pb-$6p_z$. Hence a normal band order appears, in analogy to the case of bulk PbTe. It is noted that this is different from SnTe films, where with a large number of five it becomes nontrivial phase [24].

The appearance of band inversion at two equivalent *k* points and the fact of this band inversion is driven by SOC are indicative of nontrivial TQSs. In order to provide further support, we explore the evolution of the band gap as a function of scaling factor $\lambda$ in front of the SOC term, as shown in Fig. 4(a). Obviously, it is topologically trivial in the atomic limit. With the increasing strength of SOC, the band gap closes and then reopens with the opposite band character, *i.e.*, the parity as well as the main contributing weight of the constituent atoms are reversed for CBM and VBM. These indicate a topological phase transition from a topologically trivial to nontrivial phase. More interestingly, the nontrivial band gap at the X point reaches 0.43 eV, which is larger than those of other reported 2D films [32,33,35]. The comparatively large nontrivial gap is beneficial for the future experimental preparation and makes it highly adaptable in various application environments.

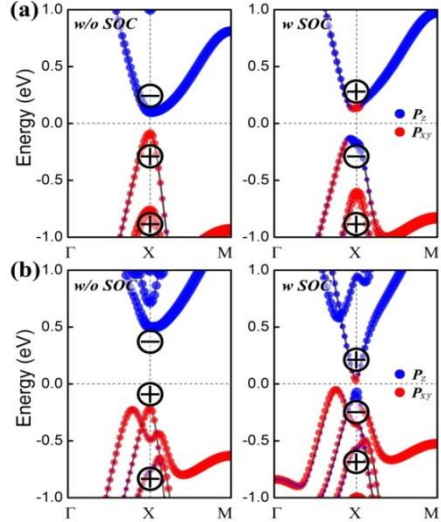

**Fig. 3** Orbital-resolved band structures for (001) (a) monolayer and (b) trilayer, with and without considering SOC respectively, weighted with the Te-$p_{x,y}$ and Pb-$p_z$ characteristics.

A hallmark of 2D TCI is the presence of two pairs of counter-propagating edge states within the bulk band gap, which move in the same (opposite) direction carrying identical (opposite) mirror eigenvalues. To prove these expectations, we calculate the local density of state (LDOS) of the edges by using the Wannier 90 package [51, 52] based on maximally localized Wannier functions (MLWFs).

Figure 5(a) shows the DFT and fitted band structures, which are in well agreement with each other. As the bulk energy bands near the Fermi level are predominantly contributed by Pb-$p_z$ and Te-$p_{x,y}$ orbitals, the MLWFs can construct from atomic $p$ orbitals, while the TB parameters are determined from MLWFs overlap matrix. Figure 5(b) shows the resultant edge states of (001) monolayer. It is noted that two Dirac points cross along the X-Γ-X direction connecting from the conduction and valence bands, one before and other after band inversion momentum space X-point, confirming its nontrivial topology. Due to the bands forming the Dirac point have different eigenvalues of the mirror operation, both interback and intraback scattering are forbidden between those states, if the mirror symmetry is strictly preserved. Additionally, the mirror is connected with the spins, and thus, transport currents are also spin-polarized due to momentum-spin locking.

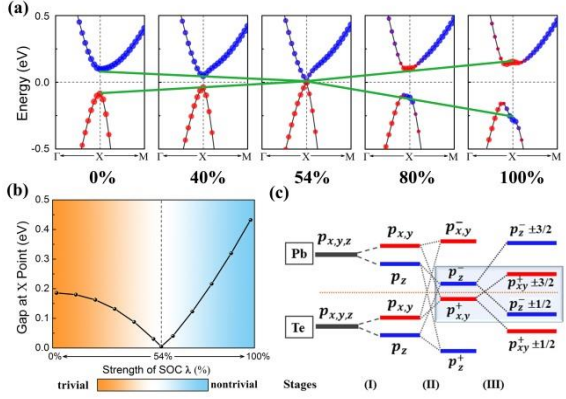

**Fig. 4** (a) Orbital-resolved band structures of the PbTe monolayer as a function of a scaling factor $\lambda$ of the SOC, while (b) show the change trend of band gap at the X point. (c) The schematic of band evolution at X point.

To elucidate the underlying origin of band inversions, we investigate the evolution of atomic orbitals of (001) monolayer. This is schematically illustrated from three stages (I), (II), and (III) in Fig. 4(c). Stage (I) indicates that the $p$-orbital are splitted into $p_{x,y}$ and $p_z$ states due to $C_4$ crystal field, pushing in-plane $p_{x,y}$ away from out-of-plane $p_z$ orbital. When forming the chemical bonds in between Pb and Te atoms, these states can further form bonding and antibonding states, labeled as $p_{xy}^-$, $p_{xy}^+$, $p_z^-$, and $p_z^+$, where the superscripts + (−) stand for even (odd) parities, as shown in stage (II). For in-plane $p_{x,y}$ orbitals, the crystal field is similar to that of bulk one where it form both σ- and π-type bonding, but σ-bond is absent for $p_z$ orbital in (001) plane. Thus, the asymmetric crystal field results in a larger onsite energy of $p_z$ than $p_{x,y}$ orbitals, thus no band inversion appears. After turning on SOC, however, the onsite energy difference is strong enough to push down $p_z$ states, leading to both Te-$p_{xy}^+$ and Pb-$p_z^-$ band order reversed, as shown in stage (III). Such a fact that indicates a nontrivial topology, can be further characterized by the mirror Chern number [53], expressed as $C_M = (C_{+i} - C_{-i})/2$, where $C_{+i}$ and $C_{-i}$ are the Chern numbers for mirror eigenvalues $+i$ and $-i$,

$$C_{\pm i} = \frac{1}{2\pi} \sum_{n<E_F} \int_{BZ} \Omega_{\pm i}(\mathbf{k}) d^2 k$$

and $\Omega_{\pm i}(\mathbf{k})$ is the Berry curvature of all occupied bands constructed from respective mirror projected states in the mirror plane, calculated according to

$$\Omega(\mathbf{k}) = -2Im \sum_{m \neq n} \frac{\langle \Psi_{n\mathbf{k}} | v_x | \Psi_{m\mathbf{k}} \rangle \langle \Psi_{m\mathbf{k}} | v_y | \Psi_{n\mathbf{k}} \rangle}{(\varepsilon_{m\mathbf{k}} - \varepsilon_{n\mathbf{k}})^2}$$

where $m$, $n$ are band indices, $\Psi_{m/n\mathbf{k}}$ and $\varepsilon_{m/n\mathbf{k}}$ are the corresponding wave functions and eigenenergies of band $m/n$, respectively, and $v_{x/y}$ are the velocity operators. The MLWFs are constructed to calculate the Berry curvature efficiently. With $z \rightarrow -z$ mirror symmetry, the calculated Chern numbers are respectively $C_{\pm i} = \pm 2$, leading to a mirror Chern number $C_M = 2$, confirming the TCI nature.

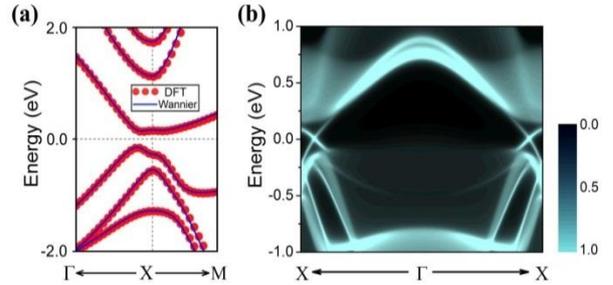

**Fig. 5** (a) Comparison of band structures of DFT (red circles) and Wannier function (blue lines) methods. (b) Local density of state (LDOS) of the edges of PbTe monolayer.

Strain plays an important role in tuning topological properties in 2D materials. Thus, we explore the strain-driven band-gap variation and TCI phase transition in (001) few-layers. The strains can be expressed as $\varepsilon_i = (a_i - a_0)/a_0$ along $i$ direction, where $a_i$ and $a_0$ are the lattice constants of 2D PbTe (001) crystals, respectively. As can be seen from Fig. 1(c), the $C_4$ symmetry makes the parity eigenvalues of the energy bands equivalent at X and Y points, preventing it from turning into a $Z_2$ TI under biaxial strains. However, the tensile strain induced topological phase transitions are found here, as illustrated in Fig. 6(a), where the variations of band gap ($E_g$, $E_g$(X)) with SOC with respect to strain are displayed. It is noted that the nontrivial topology is sensitive to $\varepsilon_i$, upon which the TCI phase transition can be tuned efficiently, i.e, when the lattice constant begins to increase, $E_g$ monotonically decreases, and even reaches zero at the X point under a critical strain of 4.8 %. Beyond this value, $E_g$ reopens up, and thus, becomes a trivial topological phase. These results are similar to the fact that bulk PbTe also becomes a 3D TCI under tensile strain [26]. Inversely, if the lattice constant are compressed, the (001) monolayer turns into a semimetal, though the $p_z$–$p_{x,y}$ band inversion is preserved. On the other hand, by using the least squares method, the data is fitted to the formula, $E_s = a\varepsilon_x^2 + b\varepsilon_x^2 + c\varepsilon_x \varepsilon_y$, where $\varepsilon_x$ and $\varepsilon_y$ are the small strains along $x$- and $y$-directions in the harmonic region. As a result of isotropy of this structure, $a$ is equal to $b$. Hence one obtains Poisson's ratio, $\nu = C_{12} / C_{11} = c / 2a = 0.13$, which is comparable with that (0.16 ) of graphene.

As is well-known, the PBE method usually underestimates the band gap of semiconductors. To obtain an accurate band structure, we performed HSE06 calculations for monolayer PbTe as an example. As shown in Fig. S2 in supplementary information, the indirect band gap is preserved with HSE06 functional, and it is enhanced to 0.67 eV, larger than the PBE value (0.19 eV). However, the band inversion character of $p_z$ and $p_{xy}$ orbitals is not affected, indicating the robustness of topological band structure on the calculated methods.

The uniaxial strain along $x$ direction, on the other hand, breaks $C_4$ symmetry and thus makes the energy bands at the $X$ and $Y$ points exhibit opposite parity eigenvalues. On this occasion, the CBM responds little to mechanical strain due to their Pb-$p_z$ character, while the energy of VBM at $X$ point changes rapidly than that at the $Y$ point, as illustrated in Fig. 6(b). Remarkably, by increasing the lattice constant along $X$ direction, the band gap at the $X$ point decreases close to zero and opens again, while the inverted band character at the $Y$ point preserves. Based on the Fu-Kane criterion [9], this scenario shows a nontrivial state. Nevertheless, when the uniaxial strain is compressed beyond 8 %, the TCI semimetal appears since the Fermi level crosses the edge of valence band in the entire 2D BZ.

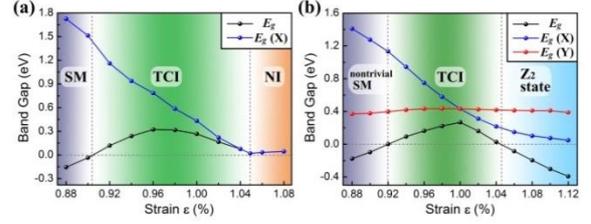

**Fig. 6** The variation of the global band gap $E_g$, band gap at $X$ point $E_g(X)$, band gap at $Y$ point $E_g(Y)$ as a function of (a) biaxial strain and (b) uniaxial strain.

Experimentally, the 2D free-standing films should be placed or grown on a substrate [54-60]. Considering the lattice constant matches well, here we select NaI as a substrate to construct a NaI/PbTe/NaI heterostructure, as shown in Fig. 7(a), where a PbTe monolayer is placed $n$-layer NaI films. Figure 7(b) presents the calculated band structures with and without SOC. One can see that a few bands appear within the band gap of NaI substrate around the Fermi level, which is still mainly contributed by PbTe monolayer. Furthermore, the $p_{x,y}$-$p_z$ band order inversion remains, and thus, its nontrivial quantum topology is preserved. On this occasion, the global gap reaches as large as 0.23 eV for NaI/PbTe/NaI heterostructure, suitable for room-temperature applications.

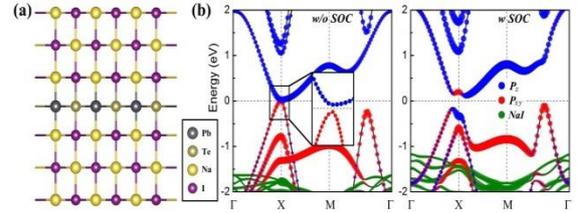

**Fig. 7** (a) Geometric structure of epitaxial growth PbTe on NaI substrate, and (b) the calculated band structure without SOC and with SOC, weighted with the Te-$p_{x,y}$, Pb-$p_z$ and NaI characteristics.

In conclusion, we through first-principles calculations to explore the electric structures and topological properties of 2D PbTe(001) few-layers, and find that it is a 2D TCI with crystalline-protected Dirac states at the edges. This nontrivial topological phase stems from the strong crystal field effect in the monolayer, which lifts the degeneracy between Te-$p_{x,y}$ and Pb-$p_z$ orbitals, resulting in a $p_{x,y}$-$p_z$ band inversion. As compared to corresponding narrow band gap in bulk PbTe, the quantum confinement leads to a larger band gap of 0.27 eV, which makes it viable for practical realization of TCI phase at room-temperature. Additionally, its nontrivial quantum topology is preserved in NaI/PbTe/NaI heterostructure. This novel 2D TCI with a large bulk gap is potential candidate in future spintronics devices with ultralow dissipation.

**Acknowledgments:** This work was supported by the National Natural Science Foundation of China (Grant No.11434006).